\documentclass[a4paper,10pt]{article}
\usepackage{graphics}
\usepackage{amsmath}
\usepackage{graphicx}
\usepackage{psfrag}
\usepackage{amssymb}
\usepackage{subfigure}
\usepackage{mathtools}
\usepackage{mathrsfs}
\usepackage{color}
\newcommand{\poo}[2]{\frac{\partial #1}{\partial #2}}
\newcommand{\po}[2]{\frac{\partial^{2} #1}{\partial #2^{2}}}
\newcommand{\p}{\partial}

\newcommand{\ve}{\varepsilon}

\title{Linear and Nonlinear Surface Waves in Electrohydrodynamics}
\author{M. Hunt, J-M. Vanden-Broeck, D. Papageorgiou, E. Parau}

\begin{document}

\maketitle

\begin{abstract}
\noindent
The problem of interest in this article are waves on a layer of finite depth governed by the Euler equations in the presence of
gravity, surface tension, and vertical electric fields. Perturbation theory is used to identify canonical
scalings and to derive a Kadomtsev-Petviashvili equation with an additional non-local term arising in interfacial
electrohydrodynamics. When the Bond number is equal to 1/3, dispersion disappears and shock waves could potentially form. In the additional limit of vanishing electric fields,
a new evolution equation is obtained which contains third- and fifth-order dispersion as well as a non-local electric field term.
\end{abstract}
\section{Introduction}
Classical water wave models leading to equations such as the Kadomtsev-Petviashvili (KP) equation have long since been used to understand the 
nonlinear phenomena and interactions of waves and are useful in the theoretical basis for further studies. The main interest here is to extend the 
results in \cite{grp1} to the three dimensional case. The importance of interfacial electrohydrodynamics phenomena has been highlighted in many 
different cases.  
\section{Set-Up and Governing Equations}
Consider a perfectly conducting, inviscid, irrotational and incompressible fluid (region 1) bounded below by a wall electrode at $y=-h$ and 
bounded above by a free surface $y=\eta (t,x,y)$, here $h$ is the mean depth of the surface. The fluid motion is described by a velocity potential 
$\varphi (t,x,y,z)$ satisfying Laplace's equation in region 1. Surface tension with coefficient $\sigma$ and gravity, $g$, are included.
The region $y>\eta (t,x,y)$, denoted by region 2, is 
occupied by a hydrodynamically passive dielectric having permittivity $\epsilon_{p}$. It is assumed that there are no free charges or currents in 
region 2 and therefore the electric field can be represented as a gradient of a potential function, $\mathbf{E}=-\nabla V$. A vertical electrical 
field is imposed by requiring that $V\sim -E_{0}y$ as $y\rightarrow\infty$, where $E_{0}$ is constant. The voltage potential satisfies the Laplace 
equation. On the free surface the Bernoulli equation holds:
\begin{equation}
\poo{\varphi}{t}+\frac{1}{2}\left[\left(\poo{\varphi}{x}\right)^{2}+\left(\poo{\varphi}{y}\right)^{2}+\left(\poo{\varphi}{z}\right)^{2}\right]
+g\eta+\frac{p}{\rho}=C
\end{equation}
The pressure $p$ is obtained through the Young-Laplace equation:
\begin{equation}
\left[\hat{\mathbf{n}}\cdot\mathbf{T}\cdot\hat{\mathbf{n}}\right]_{2}^{1}=\sigma\nabla\cdot\hat{\mathbf{n}}
\end{equation}
Where the stress tensor is given by:
\begin{equation}
\mathbf{T}=-p\delta_{ij}+\epsilon_{p}\left( E_{i}E_{j}-\frac{1}{2}\delta_{ij}E_{k}E_{k}\right)
\end{equation}
The unit normal is given by:
\begin{equation}
\hat{\mathbf{n}}=\frac{(\p_{x}\eta ,\p_{y}\eta ,-1)}{(1+(\p_{x}\eta )^{2}+(\p_{y}\eta )^{2})^{\frac{1}{2}}}
\end{equation}
The governing equations are then:
\begin{equation}
\po{\varphi}{x}+\po{\varphi}{y}+\po{\varphi}{z}=0\quad\textrm{on}\quad -h\leqslant z\leqslant\eta (t,x,y)
\end{equation}

\begin{equation}
\po{V}{x}+\po{V}{y}+\po{V}{z}=0\quad\textrm{on}\quad z>\eta (t,x,y)
\end{equation}

\begin{equation}
\poo{\eta}{t}+\poo{\varphi}{x}\poo{\eta}{x}+\poo{\varphi}{y}\poo{\eta}{y}=\poo{\varphi}{z}\quad\textrm{on}\quad z=\eta (t,x,y)
\end{equation}

\begin{equation}
\poo{\varphi}{t}+\frac{1}{2}\left[\left(\poo{\varphi}{x}\right)^{2}+\left(\poo{\varphi}{y}\right)^{2}+\left(\poo{\varphi}{z}\right)^{2}\right]
+g\eta+\frac{P}{\rho}-\frac{1}{\rho}\hat{\mathbf{n}}\cdot\Sigma\cdot\hat{\mathbf{n}}=\frac{\sigma}{\rho}\nabla\cdot\hat{\mathbf{n}}+C
\end{equation}

\begin{equation}
\poo{V}{x}+\poo{\eta}{x}\poo{V}{z}=0\quad\textrm{on}\quad z=\eta (t,x,y) 
\end{equation}

\begin{equation}
\poo{\varphi}{z}=0\quad\textrm{on}\quad z=-h
\end{equation}

\begin{equation}
V\sim -E_{0}z\quad\textrm{as}\quad z\rightarrow\infty 
\end{equation}
\section{Linear Theory}
We move into a reference frame which is moving with the fluid. This will remove the time derivatives from the governing equations.
\subsection{Finite Depth}
The scaling we use for the finite depth case is:
\begin{equation}
\begin{multlined}
x=h\hat{x},\quad y=h\hat{y},\quad z=h\hat{z},\quad \eta =h\hat{\eta} \\
t=\sqrt{\frac{\rho h^{3}}{\sigma}}\hat{t},\quad \varphi =\sqrt{\frac{\sigma h}{\rho}}\hat{\varphi},\quad V=E_{0}h\hat{V},\quad 
P=\frac{\sigma}{h}\hat{p}
\end{multlined}
\end{equation}
There are two dimensionless parameters which come out of the non-dimensionlisation process:
\begin{equation}
B=\frac{\rho h^{2}g}{\sigma},\quad E_{b}=\frac{\epsilon E_{0}^{2}\rho h}{\sigma}
\end{equation}
The parameters are called the Bond number and electric Bond number respectively. The variables are expanded as:
\begin{eqnarray}
p & = & \ve p_{1}+o(\ve ) \\
\eta & = & \ve\eta_{1}+o(\ve ) \\
\varphi & = & Ux+\ve\varphi_{1}+o(\ve ) \\
V & = & -z+\ve V_{1}+o(\ve )
\end{eqnarray}
The variables $V_{1},\eta_{1}, \varphi_{1}, p_{1}$ are written as inverse Fourier transforms:
\begin{eqnarray}
V_{1} & = & \frac{1}{(2\pi )^{2}}\int_{\mathbb{R}^{2}}e^{iklx+ily}A(k,l,z)dkdl\label{9_ch4} \\
\varphi_{1} & = & \frac{1}{(2\pi )^{2}}\int_{\mathbb{R}^{2}}e^{iklx+ily}B(k,l,z)dkdl\label{10_ch4} \\
\eta_{1} & = & \frac{1}{(2\pi )^{2}}\int_{\mathbb{R}^{2}}e^{iklx+ily}C(k,l)dkdl\label{11_ch4}
\end{eqnarray}
The expression for the free surface is then given by:
\begin{equation}\label{linear_2d}
\eta (x,y)=\frac{1}{4\pi^{2}}\int_{\mathbb{R}^{2}}\frac{\mu e^{i(kx+ly)}\hat{p}\tanh\mu}{k^{2}U^{2}-(B\mu-E_{b}\mu^{2}+\mu^{3})\tanh\mu}dkdl,
\end{equation}
where $\mu=\sqrt{k^{2}+l^{2}}$. The stability relation is given as:
\begin{equation}
U^{2}=(B\mu-E_{b}\mu^{2}+\mu^{3})\frac{\tanh\mu}{k^{2}}
\end{equation}
\begin{figure}
\centering
\includegraphics[scale=0.75]{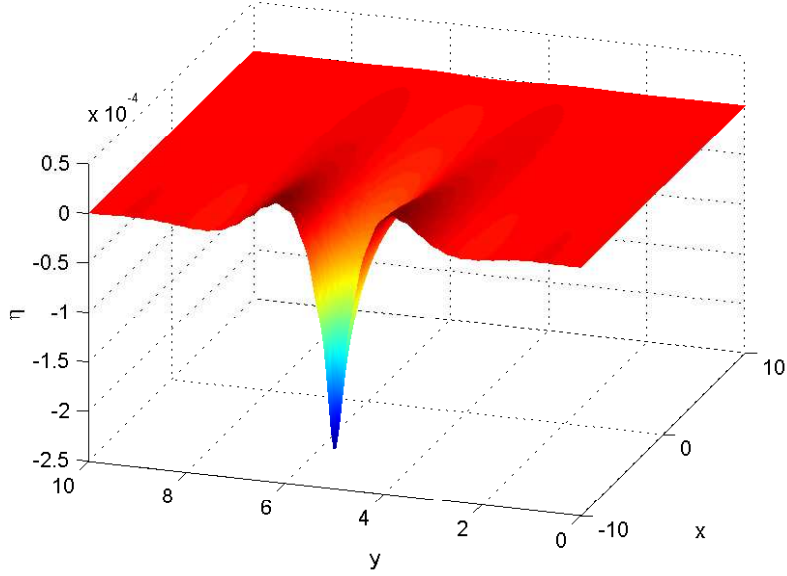}
\label{finite_profile}
\caption{Linear Waves Profiles - Finite Depth}
\end{figure}
Typical free surface profiles predicted by (\ref{linear_2d}) for $E_{b}=1.5$, $B=2$ and $U=0.5$ in figure (\ref{finite_profile}). 
\subsection{Infinite Depth}
The difference between the finite depth and the infinite depth is the boundary conditions, these new boundary conditions are given by:
\begin{eqnarray}
\poo{\varphi}{x} & \rightarrow & U\quad\textrm{as}\hspace{0.25cm}z\rightarrow\infty \\
\poo{\varphi}{z} & \rightarrow & 0\quad\textrm{as}\hspace{0.25cm}z\rightarrow -\infty
\end{eqnarray}
Using the expansion:
\begin{eqnarray}
\varphi & = & Ux+\ve\varphi_{1}+o(\ve ) \\
\eta & = & \ve\eta_{1}+o(\ve ) \\
V & = & -E_{0}z+\ve V_{1} +o(\ve ) \\
p & = & \ve p_{1} +o(\ve )
\end{eqnarray}
The same approach can be done for the infinite case, by writing the variables as inverse Fourier transforms, it is possible to write the free surface 
as:
\begin{equation}\label{id1}
\eta_{1}=\frac{1}{4\pi^{2}}\int_{\mathbb{R}^{2}}\frac{\mu\hat{p}\rho^{-1}}{k^{2}U^{2}-g\mu +\frac{\epsilon 
E_{0}^{2}}{\rho}\mu^{2}-\frac{\sigma}{\rho}\mu^{3}}e^{i(kx+ly)}dkdl
\end{equation}
Where $\hat{p}$ is the Fourier transform of the pressure. Equation (\ref{id1}) can be made clearer by the introduction of the following variables:
\begin{equation}
\alpha =U^{2}g^{-1}k,\quad \beta =U^{2}g^{-1}l,\quad \hat{x}=xgU^{-2},\quad\hat{y}=ygU^{-2},\quad\eta_{1}=g\hat{\eta}U^{-2}
\end{equation}
To obtain:
\begin{equation}
\hat{\eta}_{1}(\hat{x},\hat{y})=\frac{1}{8\pi^{2}}\int_{\mathbb{R}^{2}}\frac{\nu 
e^{-\frac{\nu^{2}}{20}}e^{i(\alpha\hat{x}+\beta\hat{y})}}{\alpha^{2}-\nu +\mu_{1}\nu^{2}-\mu_{2}\nu^{3}}d\alpha d\beta
\end{equation}
where:
\begin{equation}
\mu_{1}=\frac{\epsilon E_{0}^{2}}{\rho U^{2}}\quad\mu_{2}=\frac{\sigma g}{\rho U^{4}}
\end{equation}
\begin{figure}
\centering
\includegraphics[scale=0.75]{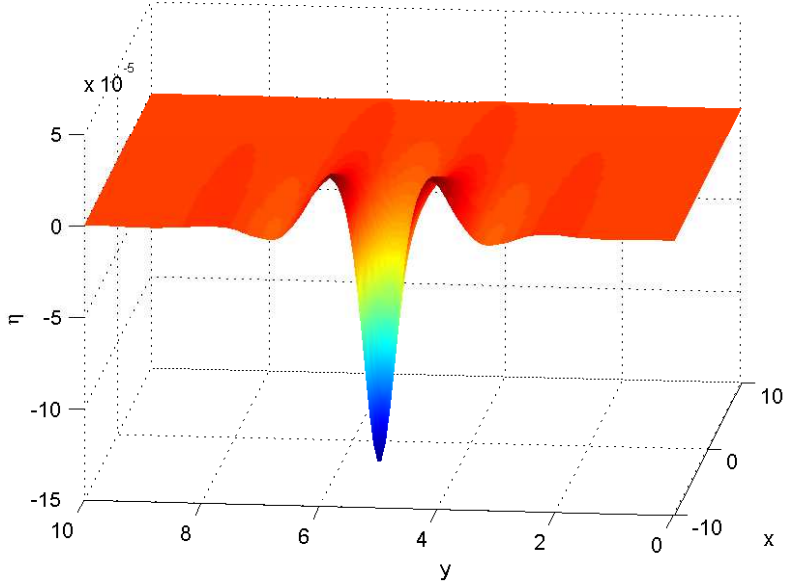}
\caption{Profile of 2D Infinite Depth Wave}
\label{infinite_2d_3}
\end{figure}
Figure (\ref{infinite_2d_3}) show the waves for infinite depth for values $\mu_{1}=1$ and $\mu_{2}=2$. When $\beta =0$, 
the denominator in the integrand can be arranged to be a perfect square provided that $\mu_{1}$ and $\mu_{2}$ satisfy a certain relation 
\begin{equation}
1+\mu_{2}=2\sqrt{\mu_{2}}.
\end{equation}
\section{Weakly Nonlinear Theory}
Previous work in this area has been carried out by Katsis and Alylas (\cite{aky1}), whereby a two fluid scenario with interface was given by $z=\eta (t,x,y)$. Analysis showed that the resulting equation was the same as that obtained in this section, despite there being a different method and different coefficients. For the weakly nonlinear theory, use the following scaling (\cite{PhD},\cite{grp1},\cite{JM4}):
\begin{equation}
x=\lambda\hat{x}, \quad y=\mu\hat{y}, \quad t=\frac{\lambda}{c_{0}}\hat{t},\quad \eta =a\hat{\eta}\quad\varphi
=\frac{g\lambda a}{c_{0}}\hat{\varphi} 
\end{equation}
\begin{equation}
V=\lambda E_{0}\hat{V}\quad z^{(1)}=h\hat{z}\quad z^{(2)}=\lambda\hat{Z},
\end{equation}
Define three parameters (\cite{grp1},\cite{PhD}):
\begin{equation}
\alpha =\frac{a}{h},\quad\beta =\frac{h^{2}}{\lambda^{2}},\quad\gamma =\frac{\lambda^{2}}{\mu^{2}}
\end{equation}
The governing equations become:
\begin{eqnarray}
\frac{1}{\beta}\po{\hat{\varphi}}{z}+\gamma\po{\hat{\varphi}}{\hat{y}}+\po{\hat{\varphi}}{\hat{x}} & = & 0\quad 
-1\leqslant\hat{z}\leqslant\alpha\hat{\eta}\label{g1} \\
\po{\hat{V}}{\hat{z}}+\gamma\po{\hat{V}}{\hat{y}}+\po{\hat{V}}{\hat{x}} & = & 0\quad \alpha\hat{\eta}\leqslant\hat{Z}\leqslant\infty\label{g2}
\end{eqnarray}

\begin{equation}
\Sigma_{33}=\frac{E_{b}}{2}\left[\left(\poo{\hat{V}}{\hat{z}}\right)^{2}-\left(\poo{\hat{V}}{x}\right)^{2}-\gamma\left(\poo{\hat{V}}{y}\right)^{2}
\right]
\end{equation}

\begin{equation}\label{V}
\poo{\hat{V}}{\hat{x}}+\alpha\sqrt{\beta}\poo{\hat{\eta}}{\hat{x}}\poo{\hat{V}}{
\hat{y}}= 0\quad\textrm{on}\hspace{0.25cm} z=\alpha\hat{\eta}. 
\end{equation}

\begin{equation}\label{free}
\frac{1}{\beta}\poo{\hat{\varphi}}{\hat{z}}=\poo{\hat{\eta}}{\hat{t}}+\alpha\poo{\hat{\varphi}}{\hat{x}}\poo{\hat{\eta}}{\hat{x}}+\alpha\gamma\poo{
\hat{\varphi}}{\hat{y}}\poo{\hat{\eta}}{\hat{y}}\quad\textrm{on}\quad z=\alpha\hat{\eta}
\end{equation}

\begin{eqnarray}
\beta B\hat{\nabla}\cdot\hat{\mathbf{n}}-\frac{E_{b}}{2\alpha} & = & \poo{\varphi}{t}+\eta 
+p+\frac{E_{b}}{\alpha}\hat{\mathbf{n}}\cdot\sigma\cdot\hat{\mathbf{n}} +{}\nonumber \\
& & 
{}+\frac{1}{2}\left[\alpha\left(\poo{\varphi}{\hat{x}}\right)^{2}+\alpha\gamma\left(\poo{\varphi}{y}\right)^{2}+\left(\poo{\varphi}{z}\right)^{2}
\right]\label{bern}
\end{eqnarray}

\begin{equation}
\frac{B\beta\left[ \p_{\hat{x}}^{2}\hat{\eta}(1+\alpha^{2}\beta\gamma (\p_{\hat{y}}\hat{\eta})^{2})+\gamma\p_{\hat{y}}^{2}\hat{\eta}(1+\alpha^{2}\beta 
(\p_{\hat{x}}\hat{\eta})^{2})-2\alpha^{2}\beta\gamma\p_{\hat{x}}\hat{\eta}\p_{\hat{y}}\hat{\eta}\p_{\hat{x}}\p_{\hat{y}}\hat{\eta}\right]}{(1+\alpha^{
2}\beta (\p_{\hat{x}}\hat{\eta})^{2}+\alpha^{2}\beta\gamma (\p_{\hat{y}}\hat{\eta})^{2})^{\frac{3}{2}}}
\end{equation}
Where:
\begin{equation}
B=\frac{\sigma}{\rho gh^{2}},\quad E_{b}=\frac{\epsilon_{p}E_{0}^{2}}{\rho gh}
\end{equation}
The non-dimensional parameter $B$ is the Bond number. Note that $E_{b}$ can be thought of as the ratio of two things, $\epsilon_{p}E_{0}^{2}/\rho$ and $gh$. The quantity $gh$ has the units of speed$^{2}$ and therefore so does $\epsilon_{p}E_{0}^{2}/\rho$. The speed $\sqrt{\epsilon_{p}E_{0}^{2}/\rho}$ is therefore characteristic to the system of interest, the quantity $\sqrt{E_{b}}$ can be thought of as the electric Foude number. The constant was calculated by using the solution $\hat{\varphi}=\hat{\eta}=0$ 
and $\hat{V}=-\hat{z}$ to show that $C=-E_{b}/2\alpha$. To begin with restrict the attention to the classical shallow water scaling by taking:
\begin{equation}\label{scale_1}
\alpha =\beta =\gamma =\varepsilon\ll 1,\quad T=\varepsilon\hat{t},\quad X=\hat{x}-\hat{t}
\end{equation}
From here on in, drop the hats. The asymptotic expansions for the variables are:
\begin{eqnarray}
\varphi & = & \varphi_{0}+\varepsilon\varphi_{1}+\varepsilon^{2}\varphi_{2}+o(\varepsilon^{2})\label{exp1} \\
\eta & = & \eta_{0}+\varepsilon\eta_{1}+o(\varepsilon )\label{exp2} \\
V & = & -z+\varepsilon^{\frac{3}{2}}V_{1}+o\left(\varepsilon^{\frac{3}{3}}\right)\label{exp3} \\
p & = & \varepsilon p_{1}\label{exp4}
\end{eqnarray}
The expansion for (\ref{exp3}) comes from examining (\ref{V}). Equation (\ref{g1}) can be solved as:
\begin{equation}
\varphi =\varphi_{0}-\ve\frac{(z+1)^{2}}{2!}\po{\varphi_{0}}{X}+\ve^{2}\left[\frac{(z+1)^{4}}{4!}\frac{\p^{4}\varphi_{0}}{\p 
X^{4}}-\frac{(z+1)^{2}}{2!}\po{\varphi_{0}}{y}\right] +o(\ve^{2})
\end{equation}
Where the $\varepsilon^{-1}$ equation shows that $\varphi_{0}=\varphi_{0}(T,X)$. The electric term in the Bernoulli equation (\ref{bern}) is:
\begin{equation}
\frac{E_{b}}{\varepsilon}\left[ -\frac{1}{2}-\varepsilon^{\frac{3}{2}}\poo{V_{1}}{Z}\right]
\end{equation}
The first part of this equation cancels the Bernoulli constant and then this leaves a term of order $\varepsilon^{\frac{3}{2}}$, so to include this 
into the order $\varepsilon$ equation, the electric Bond number is scaled according to $E_{b}=\hat{E}_{b}\sqrt{\varepsilon}$. This makes the 
$O(\varepsilon )$ Bernoulli equation become:
\begin{equation}\label{b1}
\eta_{1}=B\p_{X}^{2}-\p_{T}\varphi_{0}+\frac{1}{2}\p_{X}^{3}\varphi_{0}-p_{1}+\hat{E}_{b}\p_{z}V-\frac{1}{2}(\p_{x}\varphi_{0})^{2}
\end{equation}
Moving on to the free surface equation (\ref{free}), the $O(1)$ equation gives $\eta_{0}=\p_{X}\varphi_{0}$, which was known previously and the 
$O(\varepsilon )$ equation is:
\begin{equation}
\frac{1}{6}\p_{X}^{4}\varphi_{0}-\p_{y}^{2}\varphi_{0}-\frac{1}{2}\eta_{0}\p_{X}^{2}\varphi_{0}=\p_{T}\eta_{0}-\p_{X}\eta_{1}+\p_{X}\varphi_{0}\p_{X}
\eta_{0}
\end{equation}
Inserting the expression for $\eta_{1}$ coming from (\ref{b1}) shows that:
\begin{equation}
\frac{1}{6}\p_{X}^{4}\varphi_{0}-\p_{y}^{2}\varphi_{0}=2\p_{T}\p_{X}\varphi_{0}+\frac{1}{2}\left(\frac{1}{3}-B\right)\p_{X}^{4}\varphi_{0}+\p_{X}^{2}
p_{1}+\frac{3}{2}\p_{X}^{2}(\eta_{0}^{2})+\hat{E}_{b}\p_{X}^{2}\p_{Z}V=0
\end{equation}
In terms of $\eta_{0}$ the equation is:
\begin{equation}\label{sol1}
\poo{}{X}\left[\poo{\eta_{0}}{T}+\frac{1}{2}\left(\frac{1}{3}-B\right)\frac{\p^{3}\eta_{0}}{\p 
X^{3}}+\poo{p_{1}}{X}+\frac{3}{2}\eta_{0}\poo{\eta_{0}}{X}+\frac{\hat{E}_{b}}{2}\frac{\p^{2}V}{\p X\p Z}\right] +\frac{1}{2}\po{\eta_{0}}{y}=0
\end{equation}
The $O(1)$ equation for (\ref{g2}) needs to be solved, it is given by:
\begin{equation}\label{V_1}
\po{V_{1}}{X}+\po{V_{1}}{Z}=0
\end{equation}
This equation requires a boundary condition in order to write down a solution. All boundary conditions are zero except the one at $Z=0$. The boundary 
condition can be found by expanding (\ref{V}) to yield:
\begin{equation}
\poo{V_{1}}{X}-\poo{\eta}{X}=0
\end{equation}
Integrating this equation and assuming that the free surface dies off at infinity shows that $V_{1}=\eta$ on $Z=0$. Equation (\ref{V_1}) is solved 
by use of a Green's function. The Green's function for a 2D Laplace equation in the upper half plane is:
\begin{equation}
g(X,Z|X',Z')=\frac{1}{2\pi}\log\left[\frac{(X-X')^{2}+(Z-Z')^{2}}{(X-X')^{2}+(Z+Z')^{2}}\right]
\end{equation}
and use of Green's second identity:
\begin{equation}\label{gst}
\int_{D}g\nabla^{2}V_{1}-V_{1}\nabla^{2}gd\tau =\int_{\p D}g(\hat{\mathbf{n}}\cdot\nabla )V_{1}-V_{1}(\hat{\mathbf{n}}\cdot\nabla )gd\Sigma
\end{equation}
Inserting the boundary condition into (\ref{gst}) shows that the solution is given by:
\begin{equation}
V_{1}=-\frac{1}{\pi}\int_{\mathbb{R}}\frac{Z\eta_{0}}{(X-X')^{2}+Z^{2}}dX'
\end{equation}
Then the required result is:
\begin{equation}\label{bc1}
\poo{V_{1}}{Z}\Bigg|_{Z=0}=\mathscr{H}\left(\poo{\eta_{0}}{X}\right)
\end{equation}
So inserting (\ref{bc1}) into (\ref{sol1}) shows that:
\begin{equation}\label{sol2}
\poo{}{X}\left[\poo{\eta_{0}}{T}+\frac{1}{2}\left(\frac{1}{3}-B\right)\frac{\p^{3}\eta_{0}}{\p 
X^{3}}+\poo{p_{1}}{X}+\frac{3}{2}\eta_{0}\poo{\eta_{0}}{X}+\frac{\hat{E}_{b}}{2}\mathscr{H}\left(\po{\eta_{0}}{X}\right)\right] 
+\frac{1}{2}\po{\eta_{0}}{y}=0
\end{equation}
In terms of dimensional variables the equation is:
\begin{equation}
\begin{multlined}
 \poo{}{x}\left[\frac{1}{c_{0}}\poo{\eta}{x}+\frac{3}{2h}\eta\poo{\eta}{x}+\frac{h^{2}}{2}\left(\frac{1}{3}-B\right)\frac{\p^{3}\eta}{\p 
x^{3}}+\frac{1}{\rho g}\poo{p}{x}+\frac{E_{b}}{2}\mathscr{H}\left(\po{\eta}{x}\right)\right] + \\
+\frac{1}{2}\po{\eta}{y}=0\label{sol3}
\end{multlined}
\end{equation}
When $p=E_{b}=0$, (\ref{sol3}) reduces to the standard KP equation. In the 2D case, it has been shown that if $B<1/3$, then ``generalised'' 
solitary waves are possible. The equation which we have derived is exactly the same as equation (2.7) in \cite{aky3} but with $\hat{E}_{b}$ replaced 
with the ratio of densities of the fluids. So there appears to be a link with interfacial flows, at least on the level of equations of motion. There 
is different behaviour around $B=1/3$ and the next section details the derivation of a new equation in this case. Note that a complete analytical derivation of all terms in the equation has been given whereas they simply add the nonlinear term in \cite{aky3}.
The dispersion relation can be easily derived for the problem, it is given by:
\begin{equation}\label{disp}
\omega^{2}=\mu\left( g-\frac{\epsilon E_{0}^{2}}{\rho}\mu+\frac{\sigma}{\rho}\mu^{2}\right)\tanh h\mu
\end{equation}
where $\mu =\sqrt{k^{2}+l^{2}}$ and $k$ and $l$ are the wavenumbers. Equation (\ref{disp}) can be expanded in powers of $k$ and $l$ to obtain the 
linear portion of (\ref{sol3}).
\section{Analysis around $B=1/3$}
In this section, the scaling in (\ref{scale_1}) is replaced by:
\begin{equation}
\alpha =\varepsilon^{2},\quad\beta =\varepsilon ,\quad\gamma =\varepsilon^{2},\quad T=\varepsilon^{2}\hat{t},\quad X=\hat{x}-\hat{t}
\end{equation}
The expansions are now given by:
\begin{eqnarray}
\varphi & = & \varphi_{0}+\varepsilon\varphi_{1}+\varepsilon^{2}\varphi_{2}+\varepsilon^{3}+o(\varepsilon^{3}) \\
\eta & = & \eta_{0}+\varepsilon\eta_{1}+\varepsilon^{2}\eta_{2}+o(\varepsilon^{2}) \\
V & = & -Z+\varepsilon^{\frac{5}{2}}V_{1}+o\left(\varepsilon^{\frac{5}{2}}\right) \\
p & = & \varepsilon^{2}p_{1}+o(\varepsilon^{2}) \\
B & = & \frac{1}{3}+\varepsilon B_{1}+o(\varepsilon )
\end{eqnarray}
The governing equation for $\varphi$ is solved in exactly the same fashion as in the previous case, the solution is given by:
\begin{equation}\label{phi_1}
\varphi =\varphi_{0}(T,X)-\varepsilon\frac{(z+1)^{2}}{2!}\po{\varphi_{0}}{X}+\varepsilon^{2}\frac{(z+1)^{4}}{4!}\frac{\p^{4}\varphi_{0}}{\p 
X^{4}} 
\end{equation}
Only the derivative of the third term needs to be calculated, this is:
\begin{equation}
\poo{\varphi_{3}}{z}=-\frac{1}{5!}\frac{\p^{6}\varphi_{0}}{\p X^{6}}-\po{\varphi_{0}}{y}
\end{equation}
The $O(1)$ Bernoulli equation (\ref{bern}) once again shows that $\p_{X}\varphi_{0}=\eta_{0}$, the $O(\varepsilon )$ part of (\ref{bern}) is:
\begin{equation}
\frac{1}{3}\po{\eta_{0}}{X}=-\poo{\varphi_{1}}{X}+\eta_{1}
\end{equation}
The term $\varphi_{1}$ has been calculated in (\ref{phi_1}), the $O(\varepsilon^{2})$ part of the free surface equation (\ref{free}) is:
\begin{equation}\label{free_2}
\p_{T}\eta_{0}-\p_{X}\eta_{2}+\p_{X}\varphi_{0}\p_{X}\eta_{0}=-\eta_{0}\p_{X}^{2}\varphi_{0}+\p_{z}\varphi_{3}
\end{equation}
To find $\eta_{2}$, the $O(\varepsilon^{2})$ Bernoulli equation (\ref{bern}) is used. The same rescaling as before is used with the electric 
Bond number, $E_{b}$ as $E_{b}=\hat{E}_{b}\varepsilon^{\frac{3}{2}}$ in order to include the electric term in. The equation becomes:
\begin{equation}
B_{1}\p_{X}^{2}\eta_{0}+\frac{1}{3}\p_{X}^{2}\eta_{1}=\p_{T}\varphi_{0}-\p_{X}\varphi_{2}+\eta_{2}+p_{1}+\frac{1}{2}(\p_{X}\varphi_{0})^{2}-\hat{E}_{b
} \p_ {Z}V_{1}
\end{equation}
Inserting this into (\ref{free_2}) shows that:
\begin{equation}
\begin{multlined}
2\p_{T}\eta_{0}-B_{1}\p_{X}^{3}\eta_{0}+3\eta_{0}\p_{X}\eta_{0} +\frac{1}{45}\p_{X}^{5}\eta_{0}+\\
+\p_{y}^{2}\varphi_{0}+\p_{X}p_{1}+\hat{E}_{b}\p_{X}\p_{z}V_{1}=0
\end{multlined}
\end{equation}
In the previous section, the term $\p_{X}\p_{y}V_{1}$ was examined before and the result is just quoted, $\p_{X}\p_{y}V_{1}=\mathscr{H}(\p_{X}^{2}\eta_{0})$, 
making the final equation:
\begin{equation}
\begin{multlined}
 \poo{}{X}\left[\poo{\eta_{0}}{T}+\frac{1}{90}\frac{\p^{5}\eta_{0}}{\p 
X^{5}}+\frac{3}{2}\eta_{0}\poo{\eta_{0}}{X}-\frac{B_{1}}{2}\frac{\p^{3}\eta_{0}}{\p X^{3}}+\right. \\
\left. +\frac{\hat{E}_{b}}{2}\mathscr{H}\left(\po{\eta_{0}}{X}\right) +\frac{1}{2}\poo{p_{1}}{X}\right] +\frac{1}{2}\po{\eta_{0}}{y}=0
\end{multlined}
\end{equation}
In dimensional variables the equation becomes:
\begin{equation}\label{end_eq}
\begin{multlined}
 \poo{}{x}\left[\frac{1}{c_{0}}\poo{\eta}{t}+\poo{\eta}{x}+\frac{h^{4}}{90}\frac{\p^{5}\eta}{\p 
x^{5}}+\frac{3}{2h}\eta\poo{\eta}{x}-\frac{h^{2}}{2}\left( B-\frac{1}{3}\right)\frac{\p^{3}\eta}{\p x^{3}}+\right. \\
\left. +\frac{E_{b}h}{2}\mathscr{H}\left(\po{\eta}{x}\right) +\frac{1}{2\rho g}\poo{p}{x}\right] +\frac{1}{2}\po{\eta}{y}=0
\end{multlined}
\end{equation}

\section{Fully Nonlinear Results in Infinite Depth}

Only the infinite depth case in considered for simplicity, but an extension to include the finite depth is possible.
The conditions at infinity are
$$V\sim -E_0z,\quad \mbox{ for } z\to \infty.$$
$$\frac{\partial \phi}{\partial x}\to 1,\quad \mbox{ for } z\to -\infty.$$
The interested is in steady waves travelling with a constant speed $U$, so
choose a frame moving with the wave and non-dimensionalise the equations by using the unit velocity $U$ and the unit length
$L=\displaystyle\frac{\sigma}{\rho U^2}$. Also $V$ is non-dimensionalised 
using $E_0 L$.
  The non-dimensional 
parameters are
$$\hat{E_b}=\frac{\epsilon E_0^2}{\rho U^2} \quad \alpha=\frac{\sigma g}{\rho U^4}.$$
These are the exact same parameters obtained in the linear case with $\mu_1=\hat{E_b}$ and $\mu_2=\alpha$. This is very useful as it can be used to determine the rage of validity of the linear solution.

Set $V=0$ at the free surface $z=\eta(x,y)$ and
introduce  $\hat{V}=V+V_0 z$, which satisfies $\hat{V}\to 0$ as $z\to \infty$.
The
Bernoulli's equation becomes in this steady frame
\begin{equation}
\frac{1}{2}\left[\left(\frac{\partial \phi}{\partial x}\right)^2
+\left(\frac{\partial \phi}{\partial y}\right)^2+\left(\frac{\partial \phi}{\partial z}\right)^2
\right]-\frac{1}{2}+\alpha\eta
-K-\frac{\hat{E_b}}{2} \left(\frac{\partial \hat{V}}{\partial \hat{{\bf n}}}+\frac{1}{\sqrt{1+\eta_x^2+\eta_y^2}}
\right)^2+\frac{\hat{E_b}}{2}=0,
\label{Bern}
\end{equation}
where $\hat{{\bf n}}$ is the {\it downwards} unit normal (see equation (4)), and
$K$ is the curvature given by
$$K=\left[\frac{\eta_x}{\sqrt{1+\eta_x^2+\eta_y^2}}\right]_x+
\left[\frac{\eta_y}{\sqrt{1+\eta_x^2+\eta_y^2}}\right]_y.$$

By applying the second Greens identity for $\phi(x,y,\zeta(x,y)-x$ in the region $z<\eta(x,y)$
and for $\hat{V}$ in the region  $z>\eta(x,y)$, we obtain the boundary integral equations 
\begin{equation}
2\pi (\phi(Q)-x_Q)=\int\int_{S_F} (\phi(P)-x_P)\frac{\partial}{\partial \hat{\bf n}}\cdot
\nabla \left(\frac{1}{R_{PQ}}
\right) dS_P-\int\int_{S_F} \frac{1}{R_{PQ}} \frac{\partial}{\partial \hat{\bf n}} (\phi(P)-x_P)dS_P,
\label{bie1}
\end{equation}
\begin{equation}
2\pi \eta(x_Q,y_Q)=-\int\int_{S_F} \eta(x_P,y_P)\frac{\partial}{\partial \hat{\bf n}}
\cdot \nabla \left(\frac{1}{R_{PQ}}
\right) dS_P+\int\int_{S_F} \frac{1}{R_{PQ}}  \frac{\partial {\hat V}}
{\partial {\bf n}}dS_P,
\label{bie2}
\end{equation}
where $P$ and $Q$ are points on the free-surface $S_F$ with coordinates
$(x_P,y_P,\zeta(x_P,y_P))$ and $(x_Q,y_Q,\zeta(x_Q,y_Q))$ and 
$$R_{PQ}=|PQ|=
\sqrt{(x_P-x_Q)^2+(y_P-y_Q)^2+(\zeta(x_P,y_P)-\zeta(x_Q,y_Q))^2}.$$
The equations  (\ref{Bern})-(\ref{bie2}) are desingularized and projected on the $Oxy$ plane, and then integrated numerically
(see  Landweber \& Macagno 1969,
Forbes 1989, P\u ar\u au \& Vanden-Broeck 2002 for details). For most of the computations $40$ grid points are chosen in $x$ direction and $40$ points in $y$ direction, with a grid interval
of $dx=0.8$ and $dy=0.8$. 
The fully-localised solitary waves near the minimum of the dispersion 
relation have been computed which corresponds to the region
$$\alpha\ge \left(\frac{1+\hat{E_b}}{2}\right)^2.$$
It should be mentioned that for $\hat{E_b}=0$ this minimum corresponds to $\alpha=1/4$, which 
is the critical point for gravity-capillary waves (see P\u ar\u au {et al.} 2005).

\begin{figure}[!h]
\begin{center}
\includegraphics[scale=0.4]{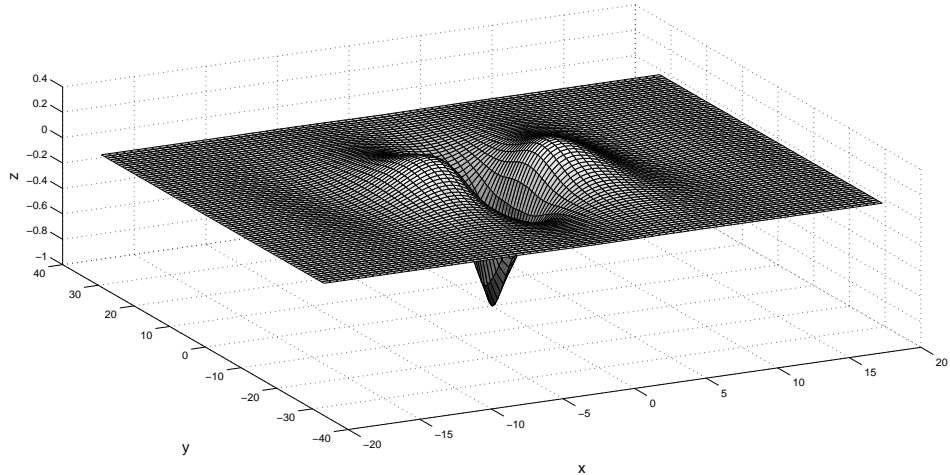}
\caption{Fully Nonlinear Infinite Depth}
\label{nonlinear_fig}
\end{center}
\end{figure}

A typical profile is given in figure \ref{nonlinear_fig}.

\section{Discussion and Conclusions}
In this paper an expression for the free surface of a perfectly conducting fluid under the influence of a vertical electrical field in the 
linear and weakly nonlinear cases has been derived in a simple, systematic way. It has been shown that, in general, to be an extension of the results for the simple 
Kadomtsev-Petviashvili (KP) equation and that this seems to be the generic expression for almost 2D equations for waves. A new equation has been derived 
(\ref{end_eq}) which incorporates a number of different physical phenomena, surface tension, electric fields and when the Bond number is close to 
$1/3$. Although the equations which have been derived here have appeared elsewhere (\cite{aky3}, \cite{pau}) with either different coefficients (as in 
\cite{aky3}) or with some terms absent (as in \cite{pau}) a simple derivation of the results has been obtained whereas in other works special assumptions have 
to be made with the coefficients (\cite{pau}) whereas the coefficients come out naturally in the analysis presented here.


\begin{thebibliography}{99}

\bibitem{PhD}
Matthew Hunt
\emph{Linear and Nonlinear Free Surface Flows in Electrohydrodynamics}
PhD Thesis, University of London

\bibitem{vb1}
Jean-Marc Vanden-Broeck,
\emph{Gravity-Capillary Free-Surface Flows}.
CUP
2010.

\bibitem{ma}
M.J. Ablowitz,
\emph{Nonlinear Dispersive Waves - Asymptotic Analysis and Solitons}
CUP
2011

\bibitem{mel1}
J.R. Melcher,
\emph{Field Coupled Surface Waves}
MIT Press
1963

\bibitem{JM1}
J-M, Vanden-Boeck, D.T. Papageorgiou
\emph{Large-amplitude capillary waves in electrified fluid sheets}
J. Fluid Mech. (2004), vol. 508

\bibitem{JM2}
E. Parau, J-M Vanden-Broeck
\emph{Nonlinear two and three dimensional free surface flows
due to moving disturbances}
Preprint

\bibitem{JM4}
J.K. Hunter, J-M. Vanden-Broeck
\emph{Solitary and periodic gravity-capillary waves of
finite amplitude}
J . Fluid Mech. (1983),vol. 134,

\bibitem{grp1}
H. Gleeson, P. Hammerton,D. T. Papageorgiou, J-M. Vanden-Broeck
\emph{A new application of the Korteweg de Vries Benjamin-Ono equation in
interfacial electrohydrodynamics}
Physics of Fluids 19, 031703 (2007)

\bibitem{Kaw1}
T. Kawahara
\emph{Oscillatory Solitary Waves in Dispersive media}
Journal of the physical society of Japan, Vol 33, Number 1, July 1972

\bibitem{russian_1}
N. M. Zubarev
\emph{Nonlinear waves on the surface of a dielectric liquid in a strong
tangential electric field}
arXiv:physics/0410097v2

\bibitem{aky1}
C. Katsis \& T.R.Akylas
\emph{On the excitation of long nonlinear water
waves by a moving pressure distribution.
Part 2. Three-dimensional effects}
J . Fluid Mech. (1987), vol. 177

\bibitem{aky2}
T.S. Yang \& T.R. Akylas
\emph{On asymmetric gravity capillary solitary waves}
J. Fluid Mech. (1997), vol. 330

\bibitem{aky3}
Boguk Kim \& T. R. Akylas
\emph{On gravity capillary lumps. Part 2.
Two-dimensional Benjamin equation}
J. Fluid Mech. (2006), vol. 557

\bibitem{pau}
L. Paumond
\emph{Towards a rigorous derivation of the fifth order KP equation}
Mathematics and Computers in Simulation 69 (2005) 477-491

\bibitem{lm}
L. Landweber\&  M.  Macagno  
\emph{Irrotational ßow about ship forms}
Iowa Institute of Hydraulic
Research Rep. IIHR 123 (1969) 1Ð33

\bibitem{ff}
L. K. Forbes
\emph{An algorithm for 3-dimensional free-surface problems in hydrodynamics}
J. Comput. Phys. 82 (1989) 330Ð347

\bibitem{pvb}
E.I. P\u ar\u au\& J.-M. Vanden-Broeck
\emph{Nonlinear two and three dimensional free surface ßows
due to moving disturbances} 
Eur. J. Mech. B/Fluids 21 (2002), 643Ð656

\bibitem{pvbc}
E.I. P\u ar\u au, J.-M. Vanden-Broeck,  Mj. Cooker
\emph{ Nonlinear three-dimensional gravity-capillary solitary waves}
J. Fluid Mechanics, 536 (2005) 99-105.


\end{thebibliography}
\end{document}